\documentclass[aps, prl, reprint, superscriptaddress]{revtex4-2}
\usepackage[utf8]{inputenc}

\usepackage{graphicx}        
\usepackage{amssymb, amsmath} 
\usepackage{siunitx}        
\usepackage{color, xcolor}  
\usepackage{soul}         

\usepackage{tabularx}      
\usepackage{multirow}     
\usepackage{verbatim}      
\usepackage{enumerate}      

\begin{document}

    \title{Clutter-Noise-Induced Phase Segregation in Binary Active Polar Flocks}

    \author{Pradeep Kumar Yadav}
    \affiliation{Departament de Física de la Matèria Condensada,
    Universitat de Barcelona, C. Martí i Franquès 1, 08028 Barcelona, Spain}
    \affiliation{Universitat de Barcelona Institute of Complex Systems (UBICS),
    Universitat de Barcelona, 08028 Barcelona, Spain}

    \author{Ignacio Pagonabarraga}
    \affiliation{Departament de Física de la Matèria Condensada,
    Universitat de Barcelona, C. Martí i Franquès 1, 08028 Barcelona, Spain}
    \affiliation{Universitat de Barcelona Institute of Complex Systems (UBICS),
    Universitat de Barcelona, 08028 Barcelona, Spain}

\begin{abstract}
    Macroscopic phase segregation in binary active mixtures can emerge solely from clutter noise—local, composition-dependent stochasticity—without repulsion or anti-alignment torques. Using an agent-based Vicsek model, we show that clutter noise alone destabilizes the mixed state, producing multi-stripe, co-moving band, and metastable counter-propagating morphologies, as well as a stable phase-segregated polar fluid. Our hydrodynamic theory shows that clutter noise suppresses local polar order and transport efficiency, consistent with the demixing observed in the agent based model.
\end{abstract}
\maketitle

\paragraph{Introduction.}
    Active systems are composed of motile units that consume energy to drive the system out of equilibrium and perform tasks~\cite{Marchetti,ramaswamy2010mechanics,Bowick22,gompper20202020}. In nature, such systems range from cytoskeleton filaments of living cells to animal herds~\cite{ouellette_2022_collective,Reynolds_2022}. The hallmark features of these systems include the emergence of self-organizing behavior, giant number fluctuations, and long-range order arising from local interactions~\cite{narayan2007,toner98,mermin1966}. A minimalist model to capture these features was proposed three decades ago by Vicsek \textit{et al.}~\cite{Vicsek1995} [the Vicsek model (VM)], and its continuum counterpart was proposed by Toner and Tu~\cite{toner98,toner95}. Numerous variants of these models have since been developed to capture specific microscopic dynamics~\cite{bechinger2016active, chate2008, Levis2020}. However, these models focus on single-species fluids; real-world scenarios are more complex, encompassing both inhomogeneity in the environment and mixtures of multiple species~\cite{GallardoNavarro2025, Lukas2023}.

   Recent studies have shifted focus toward active systems in heterogeneous environments~\cite{paoluzzi2024flocking, Codina_2022_small_obs_polarflock, vahabli2023emergence,bechinger2016active} and multispecies mixtures~\cite{parrish1997animal, Bowick22,gu2025, zarzar2022predator, dinelli2023,You20, dalziel_2021_ecosystem,Costanzo2019_size_diffes, Menzel2012_binary_species_perpendi, Navoret2013_,Maity2023_spontaneous_demixing,Kreienkamp2024_nonreciprocal, Jose2022_bacteria_experiment, rahmani2021topological}. Within these multi-component frameworks, researchers have focused heavily on binary-species systems where intra- and inter-species interactions generically involve non-reciprocal vector alignment or repulsive forces~\cite{Heiko_23, Menzel2012_binary_species_perpendi,Kurtsen25_all_interactions_repulsive, benoit_25_weaknonreciprocal}. These interactions give rise to novel types of phases such as stable anti-parallel flocking~\cite{Heiko_23,Heiko_25, benoit_25_weaknonreciprocal}, the self-organized heterogeneous aligned and clustered order phase~\cite{Oki2026_antialign_species}, chasing clusters~\cite{Kreienkamp2024_nonreciprocal}, and polarized patches~\cite{Kurtsen25_all_interactions_repulsive}. At the coarse-grained level, scalar field theories have also been proposed to study non-reciprocal interactions~\cite{saha20}. However, in the context of many biological systems~\cite{ Peled2021_bacteria_mixed_length, Ariel2015_noise_types}, a more universal yet unexplored interaction between species is physical disturbance~\cite{bangal2021phenotypic, partridge1980sensory}: the mere presence of another species acts as a localized source of stochasticity, which we refer to as clutter noise (CN). This disrupts local information transmission and compromises alignment. 

   In this Letter, we present a minimal model where intra-species interactions are governed by Vicsek-like alignment, and inter-species interactions are solely mediated by local composition-dependent stochastic noise. We demonstrate that macroscopic order and phase segregation among species do not necessarily require unfriendly instincts~\cite{Heiko_23, Heiko_25} or repulsive forces~\cite{Kurtsen25_all_interactions_repulsive}. Instead, clutter-induced noise is sufficient to destabilize a homogeneous mixed state, leading to a rich variety of non-equilibrium morphologies, including long-lived counter-propagating lanes, co-moving bands, and a stable phase-segregated polar fluid (PSPF). Our results provide new insights into non-equilibrium phase separation in active matter by showing that the system self-organizes to minimize local clutter noise, thereby maximizing transport efficiency in noisy environments. Finally, we present a coarse-grained hydrodynamic description for the density and polarization fields of the mixture, which shows that clutter noise renormalizes the linear damping term to suppress the homogeneous steady-state polar order~\cite{toner98,ramaswamy2010mechanics}. 

\paragraph{Model.}
   We consider a system of $N$ self-propelled point particles in a two-dimensional square box of linear size $L$ with periodic boundary conditions. The system consists of two species, $s\in \{a,b\}$, with $N^{s} = N/2$ particles each. Every particle $i$ of species $s$ moves at a constant speed $v_0^{s}$ and is defined by its position ${\mathbf {r}}^{s}_i$ and orientation $\theta^{s}_i$. The equations of motion are updated at each discrete time step $\Delta t$ according to:
  \begin{align}
      \mathbf{r}^{s}_i(t + \Delta t) &= \mathbf{r}^{s}_i(t) + v_0^s {\bf e}(\theta^s_i(t + \Delta t)) \Delta t,
     \label{eq:position}\\
     \theta^{s}_i(t + \Delta t) &= \operatorname{arg} \left[ \sum_{k \in {\mathcal{N}}_i^{s}} e^{i \theta^s_k(t)} \right] + \eta \zeta_i(t) +  \Lambda_i^{s} \xi_i(t), 
     \label{eq:orientation}    
    \end{align}
   where ${\bf e} (\theta^s_i) = (\cos \theta^s_i,\sin \theta^s_i)$. The first term in Eq.~\eqref{eq:orientation} represents local alignment of same-species neighbors  $k$ in $\mathcal{N}_i^s$ (including $i$ itself) within an interaction radius $R_0$. The second term describes the intrinsic Vicsek noise of amplitude $\eta \in [0,1]$, identical for both species. The interactions between opposite species are modeled as additive, composition-dependent noise, described by the final term, where the CN amplitude $\Lambda_i^{s}$  is set instantaneously by the local species composition given by:
    \begin{align}
    \Lambda_i^s = \frac{N^{\bar{s}}_i}{N^s_i + N^{\bar{s}}_i}.
    \label{eq:clutter_factor}
    \end{align}
 Here $\bar{s}$ denotes the opposite species of $s$, while $N^s_i = |\mathcal{N}_i^s|$ and $N^{\bar{s}}_i = |\mathcal{N}_i^{\bar{s}}|$ represent the local same- and opposite-species particle counts, respectively. The stochastic variables $\zeta_i$ and $\xi_i$ are independent, zero-mean white noise terms uniformly drawn from $[-\pi, \pi]$ and $[-\pi/2, \pi/2]$, respectively. The restriction of the clutter noise domain to $\xi_i \in [-\pi/2, \pi/2]$ ensures that clutter noise acts as local path deflections rather than total directional randomization, allowing particles to keep their directional persistence.

 We perform numerical simulations with fixed parameters $v_0^s = 0.5$, $R_0 = 1$, $\Delta t = 1$ in a square box of linear size $L = 128$. The total particle number density is $\rho_0 = N/L^2$, with a symmetric species composition $N^a = N^b = N/2$ that yields $\rho^s = \rho_0/2$. We explore the role of the clutter term as a function of $\eta$ and present results for $\rho_0 = 2.4$ unless otherwise stated.
\begin{figure}
\centering   \includegraphics[width=1\columnwidth]{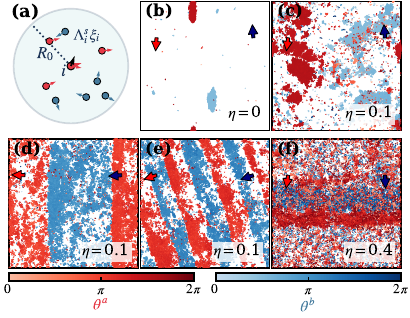}
    \caption{Emergent phases of the binary active mixture below the ordering threshold ($\eta < \eta_c$). Particle color encodes orientation via separate colorbars for each species (red: $\theta^a$; blue: $\theta^b$). Large arrows indicate the global flocking direction of each species. (a) Schematic of the alignment and clutter-noise terms setting a particle's orientation update (with simplified species coloring, distinct from the orientation colorbars in (b)--(f)). (b) Near complete phase segregation at $\eta=0$. (c)--(e) Morphologies at finite noise: (c) dynamic counter-propagating lanes with continuously refreshed interfaces, (d) macroscopic phase segregation into a single polar domain [PSPF], and (e) a kinetically trapped multi-stripe state stabilized by interfacial pinning. (f) Co-moving adjoined polar bands mutually shielding their interfaces within a disordered, gaseous background, near $\eta_c$.}
\label{fig:f1}
\end{figure}
\paragraph{Emergent Morphologies.}
    The CN amplitude, $\Lambda_i^{s}$,  a defining feature of the model, emerges locally and depends on the instantaneous local fraction of the opposite species. Hence, it acts as a random, localized interference within a crowded environement and cannot be controlled externally. Therefore, the primary control parameter we vary is the intrinsic noise $\eta$, mirroring the single-species VM. 
    

    In a single-species VM, the system remains disordered for $\eta>\eta_c$ and transitions to a polar ordered state for $\eta<\eta_c$.  While a similar ordering transition occurs in our binary mixture as $\eta$ drops below $\eta_c$, the resulting ordered phase is not a uniformly mixed polar fluid but a PSPF.
\begin{figure}[t]
    \centering   \includegraphics[width=1\columnwidth]{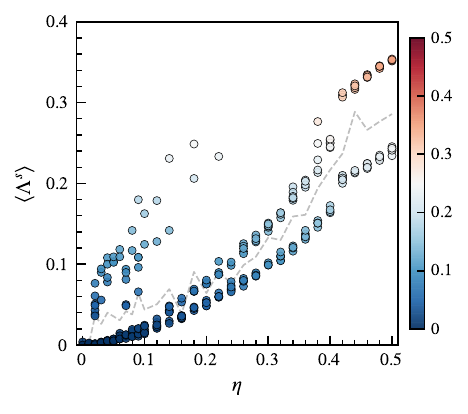}
    \caption{Average clutter-noise amplitude $\left<\Lambda^s\right>$ as a function of $\eta$, computed over 20 independent initial configurations at fixed density $\rho_0 =2.4$. The system exhibits a higher $\left<\Lambda^s\right>$ for counter-propagating lanes, and comparably lower values for both the multi-stripe and single-domain PSPF states. As $\eta \to \eta_c$, $\left<\Lambda^s\right>$ increases monotonically, consistent with the loss of phase segregation as the system approaches the disordered regime.}
    \label{fig:f2}
\end{figure}
   The underlying microscopic mechanism is illustrated in Fig.~\ref{fig:f1}(a): a particle $i$ of species $s$ (red) aligns with the mean heading of its same-species neighbors. The presence of opposite-species particles $\bar{s}$ (blue) within its interaction radius generates a clutter noise of magnitude $\Lambda^s_i$, which randomizes the particle's heading (black arrow) relative to the purely alignment-driven direction. In the absence of opposite-species neighbors, $\Lambda^s_i = 0$; the particle follows the alignment direction alone (red arrow). 
   
   For $\eta \simeq \eta_c$, the spatially homogeneous mixed isotropic state becomes unstable. In a fully mixed configuration, the local opposite-species fraction is finite, giving a CN of finite magnitude $\Lambda^s_i>0$ that disrupts intra-species alignment. Below this threshold, alignment interactions become strong enough to sustain and amplify local density fluctuations that increase the concentration of a single species and reduce $ \Lambda^{{s}}_i$ locally by sweeping out opposite-species particles. This strengthens alignment in that region, which in turn attracts further same-species particles. This creates a positive feedback  that amplifies the alignment for same-species particles and suppresses clutter noise by demixing, ultimately driving phase segregation. 
    
    This feedback yields a rich variety of phase-segregated morphologies as shown in Figs.~\ref{fig:f1}(b)--\ref{fig:f1}(f). In the absence of intrinsic noise, $\eta = 0$, the system exhibits complete phase segregation [Fig.~\ref{fig:f1}(b)], where strong alignment interactions drive both species into a polar flocking phase that remains segregated due to CN. This shows that the system selects states in which the clutter noise amplitude $\Lambda_i^s \to  0$. For finite intrinsic noise ($\eta = 0.10$), the system settles into one of three states: counter-propagating lanes [Fig.~\ref{fig:f1}(c)], a PSPF state [Fig.~\ref{fig:f1}(d)], or a multi-stripe PSPF state [Fig.~\ref{fig:f1}(e)].

    In phase-segregated counter-propagating lanes [Fig.~\ref{fig:f1}(c)], local demixing reduces $\Lambda^{s}_i$ within each domain, but the relative motion between domains continuously refreshes the interfacial composition, sustaining localized clutter fluctuations.  These local boundary fluctuations can occasionally penetrate into the bulk of a domain, reorganizing the local headings and triggering a transition into a fully phase-segregated polar flocking state [Fig.~\ref{fig:f1}(d)]. Because this bulk penetration relies on localized clutter noise variations, the macroscopic transition from lanes to a single domain is completely stochastic.

    In contrast, a multi-stripe state [Fig.~\ref{fig:f1}(e)] arises when the system is kinetically trapped, and clutter noise acts as an interfacial pinning mechanism that arrests global coarsening. For two same-species domains to coalesce, they must cross the intervening domain of the opposite species. Clutter noise at this interface disrupts the alignment of the leading particles, effectively pushing them back and preventing coalescence. As a result, the system remains trapped in a multi-stripe state characterized by sharp interfaces and local accumulation at the domain fronts. While this state may ultimately coarsen on longer runs, its coexistence alongside the single macrodomain [Fig.~\ref{fig:f1}(d)] at identical parameters highlights a highly metastable landscape where interfacial pinning halts relaxation throughout accessible simulation times.
    
   Finally, near the onset of ordering, corresponding to the microphase-separated band regime of the single-species VM, we observe a novel adjoined banding phase in which both species travel together as a single composite band through a disordered, gaseous background [Fig.~\ref{fig:f1}(f)]. Independent single-species domains would be subjected to intense interfacial clutter at both their leading and trailing edges. By co-moving in the same direction, the two species maintain a negligible relative velocity, $v_r \approx 0$ (defined below). This prevents either species from overtaking the other, so the shared interface between them stays fixed rather than being continually refreshed. As a result, local clutter at this interface is suppressed, preserving polar order strictly within the traveling band while the surrounding background remains gaseous and disordered.

\begin{figure}
    \centering   \includegraphics[width=1\columnwidth]{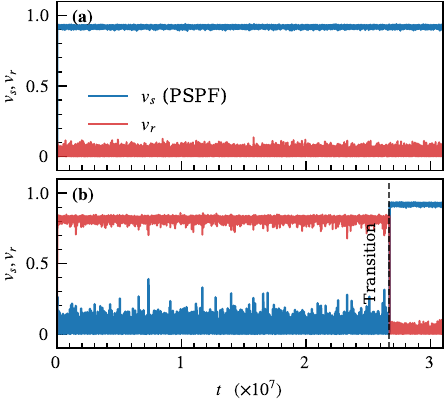}
\caption{Time evolution of the sum velocity $v_s$ and difference velocity $v_r$ at $\eta = 0.10$ and $\rho_0 = 2.4$. (a) Multi-stripe state: all domains of both species move in the same direction, so $v_s \to 1$ and $v_r \approx 0$, and both remain on stable plateaus for the entire run. (b) Counter-propagating lanes: $v_s \approx 0$ with large fluctuations, followed by a stochastic transition to the PSPF state.}
    \label{fig:f3}
\end{figure}
   To quantify the impact of clutter noise as a function of $\eta$, we compute the space- and time-averaged clutter noise $\langle\Lambda^s\rangle = \langle\Lambda_i^s(t)\rangle_{i,s,t}$. In the disordered state, $\eta > \eta_c$, the system remains homogeneously mixed.  Due to the self-inclusion of the focal particle ($N_i^s \ge 1$), a Poisson averaging of Eq.~\eqref{eq:clutter_factor} over a symmetric random distribution yields a theoretical expectation of $\langle\Lambda^s\rangle \approx 0.43$ for $\rho_0 = 2.4$. Below $\eta_c$, the system splits into two distinct branches [Fig.~\ref{fig:f2}]. Within each branch, $\langle\Lambda^s\rangle$ decreases monotonically with decreasing $\eta$, indicating that enhanced intra-species alignment sharpens spatial segregation and purges local neighborhoods of opposite-species particles. 
   
  The steady-state value of $\langle\Lambda^s\rangle$ at a given $\eta$ directly reflects the underlying morphology. Both the multi-stripe and fully phase-segregated states exhibit nearly identical values [lower branch in Fig.~\ref{fig:f2}], indicating that the degree of local mixing at the domain interfaces is independent of the specific segregated morphology. In contrast, configurations that settle into counter-propagating lanes show systematically higher $\left<\Lambda^s\right>$ [upper branch in Fig.~\ref{fig:f2}], consistent with the persistently refreshed interfacial mixing described earlier for this state.

   To quantitatively distinguish the dynamical stability of the emergent morphologies [Figs.~\ref{fig:f1}(c) and \ref{fig:f1}(e)], we define two collective order parameters: the total polar order $v_s = |\mathbf{v}^a + \mathbf{v}^b|/2$ and the relative velocity magnitude $v_r = |\mathbf{v}^a - \mathbf{v}^b|/2$, where $\mathbf{v}^s = \frac{1}{N^s} \sum_{i=1}^{N^s} \mathbf{e}(\theta_i^s)$ is the global polarization vector of species $s \in \{a,b\}$. For a co-propagating state (a PSPF or a multi-stripe state)   $\eta =0$, $v_s \approx 1$ and $v_r \approx 0$, whereas a counter-propagating state yields $v_s \approx 0$ and $v_r \approx 1$.

   The time evolution of $v_r$ and $v_s$ for a multi-stripe state and  for a counter-propagating lane is shown in Figs.~\ref{fig:f3}(a) and~\ref{fig:f3}(b), respectively. For a co-propagating state (a PSPF or a multi-stripe state), both order parameters exhibit long-lived plateaus. This shows that the multi-stripe configuration remains dynamically arrested, demonstrating that interface-localized clutter noise effectively suppresses global coarsening dynamics over a full simulation time of $3\times10^{7}$ time steps. In contrast, counter-propagating lanes exhibit pronounced temporal fluctuations in both $v_s$ and $v_r$ before undergoing a sudden, noise-induced stochastic transition to the stable PSPF state, confirming the intrinsic metastability of the counter-propagating state.

\begin{figure}
       \centering 
       \includegraphics[width=1\columnwidth]{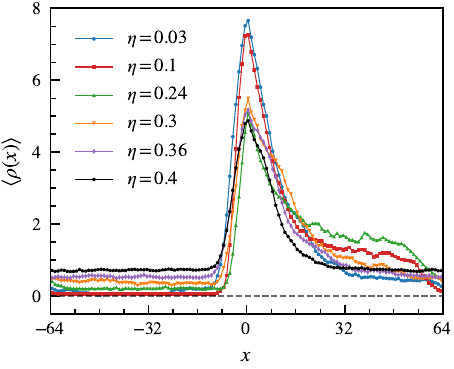}
         \caption{Density profile $\langle \rho(x) \rangle$ of a PSPF state at $\rho_0 = 2.4$, computed in a co-moving frame aligned with the domain's propagation axis and averaged over the transverse direction, for the values of $\eta$ indicated in the legend.}
         \label{fig:f4}
    \end{figure}

    To characterize the spatial profile of the PSPF state as a function of $\eta$, we isolate configurations exhibiting complete phase segregation. For these states, we compute the local density profile $\langle \rho(x) \rangle = \langle \rho^s(x) \rangle_{s,t}$ within a co-moving reference frame aligned with the domain's propagation axis ($x$), averaged over the transverse direction. This profile is plotted in Fig.~\ref{fig:f4} for various values of $\eta$ below $\eta_c$. The density exhibits a sharp peak at the leading front of the domain, followed by a gradual decay that saturates to a baseline value determined by $\eta$.  This asymmetric profile is a direct structural signature of clutter-maintained interfaces. 
    
    For $\eta \to 0$, the density essentially vanishes within the region dominated by the opposite species, whereas for $\eta \to \eta_c^-$ it saturates to a finite value, sharing qualitative structural features with single-species Vicsek bands~\cite{Vicsek1995}. As $\eta$ increases, the decay becomes less steep because intrinsic noise lets more of the opposite species leak into the domain, so the segregation weakens progressively and vanishes at $\eta_c$, following the same trend as the polar order in the single-species Vicsek model.

    To test the robustness of the segregated state, we start from phase-segregated configurations. The dependence on density is shown in Fig.~\ref{fig:system_sizes}(a): for $\rho_0 \gtrsim 1$ the profiles are nearly independent of $\rho_0$, with well-developed segregated domains, whereas at $\rho_0 = 1/10$ the system remains dilute and only weakly segregated. To test whether clutter noise selects a characteristic length scale, which would break the domain into smaller ones in large systems, we increase the system size up to $L_x = 2048$. The composition profiles collapse when lengths are rescaled by $L_x$ [Fig.~\ref{fig:system_sizes}(b)], so the domain size grows with the box and the segregation is macroscopic.
    
\paragraph{Hydrodynamic description.}
 To elucidate the macroscopic origin of clutter-driven phases, we derive hydrodynamic continuum equations for the slow variables, the conserved scalar density $\rho^s$ and the polarization field $\mathbf{P}^s$, from the Boltzmann-Ginzburg-Landau (BGL) approach in the dilute limit~\cite{bertin2006}. This framework describes the evolution of a single-particle distribution function $f^s(\mathbf{r}, \theta, t)$ for species $s$ through a kinetic equation:
 \begin{align}
    \frac{\partial f^{\mathrm{s}}}{\partial t}  
    + v_0^{s}  \mathbf{e}(\theta) \cdot \nabla f^{\mathrm{s}}
    &=  I^{\mathrm{s}}_\mathrm{sd}[f^\mathrm{s}] + I^{s\bar s}_{\mathrm{cd}}[f^s;\Lambda^s]
    \nonumber  \\ & +I_{\mathrm{coll}}^{\mathrm{ss}}[f^\mathrm{s},f^\mathrm{ s}]. 
    \label{eq:Boltzmann_eq}
    \end{align}
Here, $I^s_{\mathrm{sd}}$ represents self-diffusion under an intrinsic noise amplitude $\eta$ with a stochastic tumbling frequency $\omega_{\mathrm{sd}}$. The inter-species clutter noise term $I^{s\bar s}_{\mathrm{cd}}[f^s;\Lambda^s]$ encodes the local physical disturbances at rate $\omega_{\mathrm{cd}}$, with an amplitude set by the local composition field $\Lambda^s$, and  $I^{ss}_{\mathrm{coll}}$ captures binary intra-species alignment collisions. The explicit mathematical expressions of these individual collision integrals are in the End Matter. The hydrodynamic equation for $\rho^s(\mathbf{r},t)$, i.e., the continuity equation, emerges from the simple integration of Eq.~\eqref{eq:Boltzmann_eq} over $\theta$, which reads:
\begin{align}
  \partial_t \rho^s + v_0^s \nabla \cdot \mathbf{P}^s = 0.
  \label{eq:continuity}
\end{align}

To obtain an evolution equation for $\mathbf{P}^s$, we expand $f^s$ in angular Fourier modes $\hat{f}^s_k(\mathbf{r},t) = \int_{-\pi}^{\pi}d\theta e^{ik\theta}f^{s}({\bf r}, \theta, t)$, and using a scaling ansatz close to flocking threshold (see the End Matter), we obtain the following hydrodynamic equation for $\mathbf{P}^s$:
\begin{equation}
\begin{aligned}
\partial_t \mathbf{P}^{\mathrm{s}} = 
&\ \alpha^s(\rho^s, \Lambda^s)\mathbf{P}^{\mathrm{s}} - \beta_s |\mathbf{P}^{\mathrm{s}}|^2 \mathbf{P}^{\mathrm{s}} - \frac{v_0^{\mathrm{s}}}{2} \boldsymbol{\nabla} \rho^{\mathrm{s}} + \kappa \boldsymbol{\nabla}^2 \mathbf{P}^{\mathrm{s}} \\
&- \lambda_1 (\mathbf{P}^{\mathrm{s}} \cdot \boldsymbol{\nabla})\mathbf{P}^{\mathrm{s}} + \lambda_2 \mathbf{P}^{\mathrm{s}}(\boldsymbol{\nabla} \cdot \mathbf{P}^{\mathrm{s}}) + \lambda_3 \boldsymbol{\nabla} |\mathbf{P}^{\mathrm{s}}|^2.
\end{aligned}
\label{eq:toner_tu_vector}
\end{equation}
The above equation contains the standard Toner-Tu terms~\cite{toner98} with a modified linear growth term, $\alpha^s$, dependent on the local species composition, i.e., 
\begin{equation}
\alpha^s(\rho^s,\Lambda^s) = -\mathcal D_1 - \mathcal C_1(\Lambda^s) - \frac{2R_0v_0^s}{\pi}\left(\frac83-4\,\mathrm{sinc}(\pi\eta)\right)\rho^s .
\label{eq:alpha_linear_def}
\end{equation}
Here, $\mathcal D_1$ and $\mathcal C_1(\Lambda^s)$ are the relaxation rates of the polarization due to intrinsic and clutter noise, respectively, $\Lambda^s=\rho^{\bar s}/(\rho^s+\rho^{\bar s})$ is the local composition field (see End Matter). The clutter rate $\mathcal C_1$ vanishes in pure domains ($\Lambda^s=0$) and increases monotonically with $\Lambda^s$.

The kinetic derivation reveals that clutter noise does not act only as a linear sink via $\alpha^s$. All nonlinear and gradient coefficients ($\kappa,\beta_s,\lambda_{1,2,3}$) are inversely proportional to the relaxation rate of the second angular mode, $\nu_2$. In a mixed state, clutter adds $\mathcal C_2(\Lambda^s)=\omega_{\mathrm{cd}}[1-\mathrm{sinc}(\pi\Lambda^s)]$ to $\nu_2$ [$\approx0.36\,\omega_{\mathrm{cd}}$ at $\Lambda^s=1/2$], a term that vanishes in pure domains ($\Lambda^s\to0$) and grows monotonically with $\Lambda^s$. In mixed regions, the larger $\nu_2$ therefore reduces both the elastic constant $\kappa$, which smooths spatial variations of $\mathbf P^s$, and the self-advection coefficient $\lambda_1$, which sets how efficiently polar order is transported. Thus, clutter noise both damps polar order through $\mathcal C_1(\Lambda^s)$ in $\alpha^s$, and weakens alignment-mediated transport through $\mathcal C_2(\Lambda^s)$ in $\nu_2$.

\paragraph{Homogeneous steady state and critical threshold.}
To characterize the onset of collective motion, we examine the linear growth rate of a spatially uniform polarization perturbation about the homogeneous, symmetric mixed state ($\mathbf{P}^{\mathrm{s}} = \mathbf{0}$, $\rho^{\mathrm{s}} = \rho_0/2$), for which  $\Lambda^{\mathrm{s}} \to \Lambda_0 = 1/2$. Setting all spatial gradients and temporal derivatives to zero in Eq.~\eqref{eq:toner_tu_vector}, spontaneous polar flocking can only emerge when the linear growth operator becomes positive, $\alpha^s(\rho^s, \Lambda_0)>0$. This defines the critical threshold density, $\rho_c$, required to trigger collective motion:
\begin{equation}
\rho_c=\frac{\pi\left[\mathcal D_1+\mathcal C_1(\Lambda_0)\right]}{2R_0v_0^s\left(4\,\mathrm{sinc}(\pi\eta)-\frac83\right)}.
\label{eq:rho_c_final}
\end{equation}
Since the factor $\mathcal C_1(\Lambda_0)=\omega_{\mathrm{cd}}\Big(1-{2\sqrt2}/{\pi}\Big)\approx0.0997\,\omega_{\mathrm{cd}} $ is strictly positive, Eq.~\eqref{eq:rho_c_final} shows that $\rho_c$ increases monotonically with $\omega_{\mathrm{cd}}$, confirming that clutter noise consistently raises the ordering threshold.
\paragraph{Conclusion.---}
In summary, we have shown that the macroscopic phase-segregation mechanism in active mixtures does not require explicit anti-alignment torques or repulsive forces~\cite{Heiko_25,benoit_25_weaknonreciprocal}. Instead, it can be driven solely by a localized source of stochasticity, clutter noise (CN). This mechanism maps onto real-world scenarios where different species physically obstruct local information transmission based on phenotype or shape~\cite{partridge1980sensory,bangal2021phenotypic}. Microscopic simulations reveal that a noise-evading positive feedback loop drives the system to dynamically self-organize, minimizing local clutter noise to maximize transport efficiency.  
The resulting stable state is a phase-segregated polar fluid (PSPF) with an asymmetric, clutter-maintained interfacial profile that has no equilibrium analogue. Additionally, the system displays a rich variety of long-lived morphologies, including frozen multi-stripe configurations, a novel co-moving adjoined banding phase, and metastable counter-propagating lanes that eventually decay stochastically into the PSPF state.

Our coarse-grained hydrodynamic equations quantify these observations. The clutter noise enters through a composition-dependent linear growth rate $\alpha^s(\rho^s, \Lambda^s)$, which reduces the local ordering tendency in mixed regions, and yields a closed-form critical threshold density $\rho_c$ that increases monotonically with the clutter frequency $\omega_{\mathrm{cd}}$. The clutter noise also renormalizes the nonlinear transport coefficients $\kappa, \beta_s, \lambda_{1,2,3}$ through the composition-dependent relaxation rate $\nu_2(\rho^s, \Lambda^s)$,  so that mixed regions have a reduced elastic constant and weaker self-advection. These continuum terms emerge naturally from the collision integrals of the kinetic theory, providing a macroscopic rationale for why local demixing is dynamically favored.

Our framework provides a mechanism for demixing in mixed-culture bacterial colonies~\cite{Peled2021_bacteria_mixed_length} and multi-component synthetic active colloids. It may also be relevant to pedestrian dynamics, suggesting how clutter-induced delays arise in crowded flows~\cite{helbing1995social}. Future extensions to asymmetric species compositions or unequal self-propulsion speeds may reveal further non-equilibrium phases in multi-species active matter.

\paragraph{Acknowledgement---}P.K.Y. acknowledges funding from the
European Commission through the Horizon Europe project MULTIXSCALE
under Grant Agreement No. 101093169. I. P. acknowledges
the support from Ministerio de Ciencia, Innovación y
Universidades MCIU/AEI/FEDER for financial support
under Grant Agreement No. PID2024-156516NB-100
AEI/FEDER-EU, and Generalitat de Catalunya for financial support under Program Icrea Acad\`emia and Project
No. 2021SGR-673. 

\bibliography{refs}
\onecolumngrid
\begin{center}
\textbf{End Matter}
\end{center}
\twocolumngrid

\appendix
\renewcommand{\thesection}{A}
\setcounter{equation}{0}
\section{Derivation of the hydrodynamic equations}
\label{sec:appendixA}
 The single particle distribution function $ f^s({\mathbf {r}}, \theta, t)$ of species $s \in \{a,b\}$ evolves according to Eq.~\eqref{eq:Boltzmann_eq}. The terms $I^{\mathrm{s}}_{\mathrm{sd}}$ and $I^{\mathrm{s\bar{s}}}_{\mathrm{cd}}$ are explicitly described as
\begin{align}
I^{\mathrm{s}}_{\mathrm{sd}}[f^\mathrm{s}] = &-\omega_{\mathrm{sd}} f^{\mathrm{s}}(\mathbf{r},\theta,t) + \omega_{\mathrm{sd}} \int_{-\pi}^{\pi} d\theta^{\prime} \int_{-\pi}^{\pi} d\zeta \, p_{\zeta}(\zeta)  \nonumber \\
& \sum_{m=-\infty}^{\infty} f^{\mathrm{s}}(\mathbf{r},\theta^{\prime},t) 
 \times \delta(\theta^{\prime} + \eta \zeta -\theta + 2m\pi),
\label{eq:I_sd}
\end{align}
\begin{align}
I^{s\bar s}_{\mathrm{cd}}[f^s;\Lambda^s] = &-\omega_{\mathrm{cd}} f^{\mathrm{s}}(\mathbf{r},\theta,t) \nonumber \\
&+ \omega_{\mathrm{cd}}\int_{-\pi}^{\pi} d\theta^{\prime} \int_{-\pi/2}^{\pi/2} d\xi \, p_{\xi}(\xi)  \nonumber \\
& \sum_{m=-\infty}^{\infty} f^{\mathrm{s}}(\mathbf{r},\theta^{\prime},t)  \times \delta\left(\theta^{\prime} + \Lambda^s \xi -\theta + 2m \pi\right),
\label{eq:I_cd}
\end{align}
where $p_{\zeta}(\zeta) = 1/(2\pi)$ and $p_{\xi}(\xi) = 1/(\pi)$ represent uniform white noise distributions over $[-\pi, \pi]$ and  $[-\pi/2, \pi/2]$, respectively.

The alignment integral $I_{\mathrm{coll}}^{\mathrm{ss}}[f^\mathrm{s}, f^\mathrm{s}]$ captures the binary intra-species alignment within a cross-section $2R_0$:
\begin{equation}
\begin{split}
I_{\mathrm{coll}}^{\mathrm{ss}}&[g^{\mathrm{s}}, h^{\mathrm{s}}] = 
-2 R_0 v_0^{\mathrm{s}} \, g^{\mathrm{s}}(\mathbf{r},\theta,t) \\
&\times \int_{-\pi}^{\pi} d\theta'\, \left| \mathbf{e}(\theta') - \mathbf{e}(\theta) \right| h^{\mathrm{s}}(\mathbf{r},\theta',t) \\
&+ 2 R_0 v_0^{\mathrm{s}} \int_{-\pi}^{\pi} d\theta_1 d\theta_2 \int_{-\pi}^{\pi} d\zeta \, p_{\zeta}(\zeta) \\
&\times \left| \mathbf{e}(\theta_2) - \mathbf{e}(\theta_1) \right| g^{\mathrm{s}}(\mathbf{r},\theta_1,t)\, h^{\mathrm{s}}(\mathbf{r},\theta_2,t) \\
&\times \sum_{m} \delta\!\left(\bar{\theta} + \eta \zeta - \theta + 2 m \pi \right),
\end{split}
\label{eq:I_coll}
\end{equation}
where $\bar{\theta} = \text{Arg}[e^{i\theta_1} + e^{i\theta_2}]$ is the mean alignment direction.
 
Further, we expand $f^s$ in angular Fourier modes using the relation ${f}^s=(2 \pi)^{-1}\sum_{k =-\infty}^{\infty} e^{-ik\theta}\hat{f}^{s}_{k}({\bf r},t)$. Multiplying Eq.~\eqref{eq:Boltzmann_eq} by $e^{ik\theta}$ and integrating over $\theta$ yields (hereinafter, we use a simplified notation of $\hat{f}^s$ by dropping explicit dependence on $\mathbf{r}$ and $t$):
\begin{equation}
\partial_t \hat{f}_k^{s} + \tfrac{v_0^{s} }{2}( \nabla^{*} \hat{f}_{k+1}^{s} + \nabla \hat{f}_{k-1}^{s} ) =
-[\mathcal{D}_k  + \mathcal{C}_k(\Lambda^s)] \hat{f}_k^{\mathrm{s}} - \mathcal{A}_k^s,
\label{eq:f_k_evolution}
\end{equation}
where the complex gradient operators are $\nabla \equiv \partial_x + i \partial_y$, $\nabla^* \equiv \partial_x - i \partial_y$. The linear relaxation rates are $\mathcal{D}_k = \omega_{sd}[1-\mathrm{sinc}(k\pi\eta)]$ and $\mathcal{C}_k(\Lambda^s) = \omega_{cd}[1-\mathrm{sinc}(\Lambda^s k \pi/2 )]$ with $\mathrm{sinc}(x)= \sin(x)/x$.  The CN acts only through $I^{s\bar s}_{\mathrm{cd}}$: for $\Lambda^s=0$ the kernel reduces to $\delta(\theta'-\theta)$ and the gain and loss terms cancel ($\mathcal{C}_k=0$). As in standard BGL treatments, the intrinsic and clutter noises, which act simultaneously in the microscopic update, are modeled as independent stochastic events.  Using the angular kernel integration rules from~\cite{bertin2006}, the non-linear alignment term reads $\mathcal{A}_k^s = \frac{2 R_0 v_0^s}{\pi} \sum_q [I_q - \mathrm{sinc}(k\pi\eta) I_{q-k/2}]\hat{f}_q^{\mathrm{s}} \hat{f}_{k-q}^{\mathrm{s}}$ where $I_q= \int_{-\pi}^{\pi}d \theta |\sin(\theta/2)|\cos(q\theta)$. The values needed below are $I_0=4$, $I_{\pm1}=-4/3$, $I_{\pm2}=-4/15$, $I_{\pm1/2}=2$, and $I_{\pm3/2}=-2$. 

 The macroscopic hydrodynamic fields correspond to the particle density $\rho^s = \hat{f}_0^s$ and the complex polarization $P^s \equiv \hat f_1^s=P_x^s + iP_y^s$, whose real and imaginary parts form the vector field $\mathbf P^s=(P_x^s,P_y^s)$; since $f^s$ is real, $\hat f_{-k}^s=(\hat f_k^s)^*$. Close to the flocking threshold we adopt the scaling ansatz~\cite{bertin2006} $\hat f_k^s\sim\mathcal O(\epsilon^{|k|})$ for $k\neq0$, $\rho^s-\rho_0^s\sim\mathcal O(\epsilon)$ for both species (with $\rho_0^s=\rho_0/2$), so that  $\nabla\sim\partial_t\sim\mathcal O(\epsilon)$.

Truncating Eq.~\eqref{eq:f_k_evolution} for $k=1$ at order $\epsilon^3$, only the terms with $q\in\{-1,0,1,2\}$ in $\mathcal{A}_1^s$ contribute, which gives:
\begin{equation}
\begin{aligned}
\frac{\partial \hat{f}_1^{\mathrm{s}}}{\partial t} 
&+ \frac{v_0^{\mathrm{s}}}{2} \left( \nabla^{*} \hat{f}_{2}^{\mathrm{s}} + \nabla \rho^{\mathrm{s}} \right) = 
- \left[ \mathcal{D}_1 + \mathcal{C}_1(\Lambda^s) \right] \hat{f}_1^{\mathrm{s}} \\
&- \frac{2 R_0 v_0^{\mathrm{s}}}{\pi} \left(\frac{8}{3} - 4 \mathrm{sinc}(\pi \eta) \right) \rho^{\mathrm{s}}\hat{f}_1^{\mathrm{s}} \\
&+ \frac{2 R_0 v_0^{\mathrm{s}}}{\pi} \left(\frac{8}{5} - 4\mathrm{sinc}(\pi \eta)\right) \hat{f}_2^{\mathrm{s}} (\hat{f}_{1}^{\mathrm{s}})^{*}.
\end{aligned}
\label{eq:f1_mode}
\end{equation}
The above equation involves $\hat{f}_2^s \sim\mathcal{O}(\epsilon^2)$. To find a closure relation for $\hat{f}_2^s$, we write Eq.~\eqref{eq:f_k_evolution} for $k=2$ up to $\mathcal{O}(\epsilon^2)$, where only $q \in \{0,1,2\}$ contribute to $\mathcal{A}_2^s$ (the terms corresponding to $q=-1,3$ are $\mathcal{O}(\epsilon^4)$):
\begin{equation}
\begin{aligned}
\partial_t \hat{f}_2^{s} &+ \tfrac{1}{2}v_0^{s} \nabla \hat{f}_{1}^{s} = - \left[\mathcal{D}_2  + \mathcal{C}_2(\Lambda^s)\right] \hat{f}_2^{\mathrm{s}} \\
&- \frac{16 R_0 v_0^s}{\pi} \left[\frac{7}{15} + \frac{1}{3}\mathrm{sinc}(2\pi\eta)\right] \rho^s \hat{f}_2^s \\
&+ \frac{8 R_0 v_0^s}{\pi}\left[\frac{1}{3} + \mathrm{sinc}(2\pi\eta)\right] (\hat{f}_1^s)^2.
\end{aligned}
\label{eq:f2_evolution}
\end{equation}
Neglecting $\partial_t\hat f_2^s\sim\mathcal O(\epsilon^3)$ and keeping terms up to $\mathcal O(\epsilon^2)$, Eq.~\eqref{eq:f2_evolution} gives
\begin{align}
\hat{f}_2^{\mathrm{s}} = \frac{1}{\nu_2(\rho^{\mathrm{s}}, \Lambda^s)} \left\{ -\frac{v_0^{\mathrm{s}}}{2} \nabla \hat{f}_{1}^{\mathrm{s}} + \frac{8 R_0 v_0^{\mathrm{s}}}{\pi}\left(\frac{1}{3} + \mathrm{sinc}(2\pi\eta) \right)(\hat{f}_1^{\mathrm{s}})^2 \right\}, \label{eq:f2_slaved_clean} \\
\nu_2(\rho^{\mathrm{s}}, \Lambda^s) = \mathcal{D}_2  + \mathcal{C}_2(\Lambda^s) + \frac{16 R_0 v_0^{\mathrm{s}}}{\pi} \left[\frac{7}{15} + \frac{1}{3}\mathrm{sinc}(2\pi\eta)\right] \rho^{\mathrm{s}}. \label{eq:nu2_definition}
\end{align}
with $\mathcal C_2(\Lambda^s)=\omega_{\mathrm{cd}}[1-\mathrm{sinc}(\pi\Lambda^s)]$. Following~\cite{bertin2006}, we keep the full $\rho^s$ in $\nu_2(\rho^{\mathrm{s}}, \Lambda^s)$; the difference from $\rho_0^s$ contributes only at $\mathcal O(\epsilon^4)$ in the polarization equation. Note that $\nu_2(\rho^{\mathrm{s}}, \Lambda^s)>0$ for all $\eta$ and $\Lambda^s$.

We plug in the resulting expression into Eq.~\eqref{eq:f1_mode} and obtain the hydrodynamic equation for $P^s$: 
\begin{widetext}
\begin{align}
\frac{\partial P^{\mathrm{s}}}{\partial t} = 
&\ \alpha^s(\rho^{\mathrm{s}}, \Lambda^{{s}}) P^{\mathrm{s}} 
+ \frac{16 R_0^2 (v_0^{\mathrm{s}})^2}{\pi^2 \nu_2(\rho^{\mathrm{s}},\Lambda^s)} \left(\frac{8}{5} - 4\mathrm{sinc}(\pi \eta)\right) \left(\frac{1}{3} + \mathrm{sinc}(2\pi\eta) \right) |P^{\mathrm{s}}|^2 P^{\mathrm{s}}
- \frac{v_0^{\mathrm{s}}}{2} \nabla \rho^{\mathrm{s}} 
+ \frac{(v_0^{\mathrm{s}})^2}{4} \nabla^{*} \left(\frac{1}{\nu_2(\rho^{\mathrm{s}},\Lambda^s)} \nabla P^{\mathrm{s}}\right) \nonumber \\
&- \frac{4 R_0 (v_0^{\mathrm{s}})^2}{\pi}\left(\frac{1}{3} + \mathrm{sinc}(2\pi\eta) \right) \nabla^{*} \left( \frac{(P^{\mathrm{s}})^2}{\nu_2(\rho^{\mathrm{s}},\Lambda^s)} \right) 
- \frac{R_0 (v_0^{\mathrm{s}})^2}{\pi \nu_2(\rho^{\mathrm{s}},\Lambda^s)} \left(\frac{8}{5} - 4\mathrm{sinc}(\pi \eta)\right) (P^{s})^{*} \nabla P^{\mathrm{s}}.
\label{eq:P_closed_toner_tu_style}
\end{align}
\end{widetext}

We write Eq.~\eqref{eq:P_closed_toner_tu_style} in vector form (Eq.\eqref{eq:toner_tu_vector}) with new coefficients $\alpha^s, \kappa, \beta_s,\lambda_1, \lambda_2, \lambda_3 $ to make it easier to read, where
\begin{align}
\kappa &= \frac{(v_0^{\mathrm{s}})^2}{4\nu_2(\rho^{\mathrm{s}}, \Lambda^s)}, \label{eq:coeff_kappa} \\
\beta_s &= -\frac{16 R_0^2 (v_0^{\mathrm{s}})^2 \left(\frac{8}{5} - 4\mathrm{sinc}(\pi \eta)\right) \left(\frac{1}{3} + \mathrm{sinc}(2\pi\eta)\right)}{\pi^2 \nu_2(\rho^{\mathrm{s}},\Lambda^s)}, \label{eq:coeff_betas}
\end{align}
\begin{align}
 \lambda_1 = \frac{R_0 (v_0^s)^2 \left[8\left(\tfrac13+\mathrm{sinc}(2\pi\eta)\right) + \left(\tfrac85-4\,\mathrm{sinc}(\pi\eta)\right)\right]} {\pi \nu_2(\rho^{\mathrm{s}},\Lambda^s)},\\
\lambda_2 = \frac{R_0 (v_0^s)^2 \left[-8\left(\tfrac13+\mathrm{sinc}(2\pi\eta)\right) + \left(\tfrac85-4\,\mathrm{sinc}(\pi\eta)\right)\right]}{\pi \nu_2(\rho^{\mathrm{s}},\Lambda^s)},\\
\lambda_3 = \frac{R_0 (v_0^s)^2 \left[4\left(\tfrac13+\mathrm{sinc}(2\pi\eta)\right) - \tfrac12\left(\tfrac85-4\,\mathrm{sinc}(\pi\eta)\right)\right]} {\pi \nu_2(\rho^{\mathrm{s}},\Lambda^s)}.   
\end{align}
Whenever an ordered state exists, i.e.\ $4\,\mathrm{sinc}(\pi\eta)>8/3$ [required for $\rho_c>0$ in Eq.~\eqref{eq:rho_c_final}], one has $\frac85-4\,\mathrm{sinc}(\pi\eta)=-\frac{16}{15}-\left[4\,\mathrm{sinc}(\pi\eta)-\frac83\right]<0$, while $\frac13+\mathrm{sinc}(2\pi\eta)>0$ for all $\eta$; hence $\beta_s>0$ and the cubic term saturates the polar order.
\begin{figure}[t]
    \centering
    \includegraphics[width=0.80\linewidth]{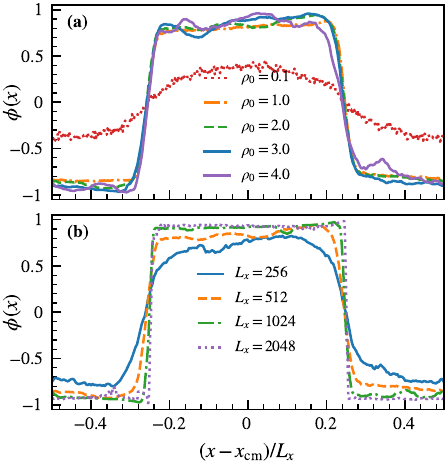}
    \caption{\label{fig:system_sizes} Steady-state profiles of the composition order parameter $\phi(x) = \frac{\rho_a(x) - \rho_b(x)}{\rho_a(x) + \rho_b(x)}$ measured in a co-moving frame centered at $x_{\mathrm{cm}}$, starting from phase-segregated configurations at $\eta = 0.10$. (a) Density sweep from $\rho_0 = 0.1$ to $4.0$ at fixed box dimensions $L_x = 512$ and $L_y = 128$. (b) Scaling collapse across system sizes from $L_x = 256$ to $2048$ at $\rho_0 = 2.0$, with fixed aspect ratio $L_y = L_x/4$. The collapse shows that the domain size grows with the system size, so clutter noise selects no characteristic length scale and the segregation is macroscopic.}
\end{figure}

\end{document}